\documentstyle[aps]{revtex}
\input epsf
\def\jepsfbox#1{\typeout{#1} \epsfbox{#1}}
\def\plottwo#1#2{\centering \leavevmode
\epsfxsize=.45\columnwidth \jepsfbox{#1} \hfil
\epsfxsize=.45\columnwidth \jepsfbox{#2}}
\def\rarrow{\rightarrow}
\def\etal{{\it et al.\ }}
\def\eg{{\it e.g.~}}

\def\rmmat#1{{\hbox{\rm #1}}}
\def\rmscr#1{\rmmat{\scriptsize #1}}
\newcommand{\be}{\begin{equation}}
\newcommand{\ee}{\end{equation}}
\newcommand{\ba}{\begin{eqnarray}}
\newcommand{\ea}{\end{eqnarray}}
\def\p{\partial}
\def\d{{\rm d}}

\def\pp#1#2{\frac{\p #1}{\p #2}}
\def\csch{ \mathop{\rm csch}\nolimits} 
\def\IM{\mathop{\rm Im}\nolimits}   
\def\RE{\mathop{\rm Re}\nolimits}   
\def\figref#1{Fig.~\ref{fig:#1}}
\def\eqref#1{Eq.~\ref{eq:#1}}
\begin{document}
\draft
\newcommand{\bfi}{{\vec B}}
\newcommand{\efi}{{\vec E}}
\newcommand{\lag}{{\cal L}}
\newcommand{\dLIII}{{\frac{\partial^3 \lag}{\partial I^3}}}
\newcommand{\dLII}{{\frac{\partial^2 \lag}{\partial I^2}}}
\newcommand{\dLI}{{\frac{\partial \lag}{\partial I}}}
\newcommand{\dLKKK}{{\frac{\partial^3 \lag}{\partial K^3}}}
\newcommand{\dLKK}{{\frac{\partial^2 \lag}{\partial K^2}}}
\newcommand{\dLK}{{\frac{\partial \lag}{\partial K}}}
\newcommand{\dLIK}{{\frac{\partial^2 \lag}{\partial I \partial K}}}
\title{An Analytic Form for the Effective Lagrangian of QED and its
Application to Pair Production and Photon Splitting}
\author{Jeremy S. Heyl \and Lars Hernquist\thanks{Presidential Faculty Fellow}}
\address{Lick Observatory, 
University of California, Santa Cruz, California 95064, USA}
\maketitle
\begin{abstract}
We derive an analytic form for the Heisenberg-Euler Lagrangian in the
limit where the component of the electric field parallel to the
magnetic field is small.  We expand these analytic functions to all
orders in the field strength ($F_{\mu\nu}F^{\mu\nu}$) in the limits of
weak and strong fields, and use these functions to estimate the
pair-production rate in arbitrarily strong electric fields and the
photon-splitting rate in arbitrarily strong magnetic fields.
\end{abstract}

\section{Introduction: The One-Loop Effective Lagrangian of QED}
When one-loop corrections are included in the Lagrangian of the
electromagnetic field one obtains a non-linear correction term:
\be
\lag = \lag_0 + \lag_1.
\label{eq:lagdef}
\ee
Both terms of the Lagrangian can be written in terms of the Lorentz
invariants,
\be
I = F_{\mu\nu} F^{\mu\nu} = 2 \left ( |\bfi|^2 - |\efi|^2 \right )
\label{eq:Idef}
\ee
and
\be
K = \{\epsilon^{\lambda\rho\mu\nu} F_{\lambda\rho} F_{\mu\nu} \}^2 =
	- (4 \efi \cdot \bfi )^2,
\label{eq:Kdef}
\ee
following Heisenberg and Euler \cite{Heis36}.  We do not expect terms
which are odd powers of
$\epsilon^{\lambda\rho\mu\nu} F_{\lambda\rho} F_{\mu\nu}$ to appear in
the effective Lagrangian as these terms would yield a Lagrangian
which would violate the $C$ and $P$ symmetries of the tree-level
Lagrangian.

Heisenberg and Euler\cite{Heis36} and Weisskopf\cite{Weis36}
independently derived the effective Lagrangian of the electromagnetic
field using electron-hole theory.  Schwinger\cite{Schw51} later
rederived the same result using quantum electrodynamics.  In
rationalized electromagnetic units, the Lagrangian is given by
\begin{eqnarray}
\lag_0 & = & -{1 \over 4} I \label{eq:lag0def} \\
\lag_1 & = & {e^2 \over h c} \int_0^\infty e^{-\zeta} 
{\d \zeta \over \zeta^3} \left \{ i \zeta^2 {\sqrt{-K} \over 4} \times
\phantom{ 
\cos \left ( {\zeta \over B_k} \sqrt{-{I\over 2} + i \sqrt{K}} \right ) 
\over
\cos \left ( {\zeta \over B_k} \sqrt{-{I\over 2} + i \sqrt{K}} \right ) } 
\right . \\*
\nonumber
& & ~~ \left . 
{ \cos \left ( {\zeta \over B_k} \sqrt{-{I\over 2} + i {\sqrt{-K}\over 2}} \right ) +
\cos \left ( {\zeta \over B_k} \sqrt{-{I\over 2} - i {\sqrt{-K}\over 2}} \right ) \over
\cos \left ( {\zeta \over B_k} \sqrt{-{I\over 2} + i {\sqrt{-K}\over 2}} \right ) -
\cos \left ( {\zeta \over B_k} \sqrt{-{I\over 2} - i {\sqrt{-K}\over 2}} \right ) } 
 + |B_k|^2 + {\zeta^2 \over 6} I \right \}.
\label{eq:lag1def}
\ea
where $B_k = E_k = {m^2 c^3 \over e \hbar} \approx 2.2 \times 10^{15}
\rmmat{\,V\,cm}^{-1} \approx 4.4 \times 10^{13}$\,G. 
Both Heisenberg and Euler\cite{Heis36}, and Mielniczuk\cite{Miel82}
present alternative
expressions for these integrals in terms of infinite series.

\section{The Analytic Expansion}

For many interesting problems, one needs an expansion of this
Lagrangian in the limit where the
component of the electric field in the direction of the magnetic field
is small (small $K$)
\be
\lag_1 = \lag_1(I,0) + K \left . \pp{\lag_1}{K} \right |_{K=0} +
\frac{K^2}{2} \left . \frac{\partial^2 \lag_1}{\partial K^2} \right
|_{K=0} + \cdots
\label{eq:lag1exp}
\ee
where the terms are given by the following integrals
\ba
\lag_1(I,0) & = & {e^2 \over h c} \frac{I}{2} X_0\left(\frac{1}{\xi}\right)
\nonumber \\*
\label{eq:x0int}
	    & = & {e^2 \over h c} \frac{I}{2} \int_0^\infty e^{-u/\xi} 
	{\d u \over u^3} \left (-u \coth u + 1 + {u^2 \over 3} \right
)
\\
 \left . \pp{\lag_1}{K} \right |_{K=0} & = & {e^2 \over h c}
\frac{1}{16 I} X_1\left(\frac{1}{\xi}\right) \nonumber \\*
\label{eq:x1int}
& = & {e^2 \over h c} \frac{1}{16 I} \int_0^\infty e^{-u/\xi} 
	{\d u \over u^2} \left (\coth u - \frac{2}{3} u^2 \coth u - u
\csch^2 u \right ) \\
 \left . \frac{\partial^2 \lag_1}{\partial K^2} \right |_{K=0} & = &
{e^2 \over h c} \frac{1}{384 I^3} X_2\left(\frac{1}{\xi}\right) \nonumber \\
& = & {e^2 \over h c} \frac{1}{384 I^3} \int_0^\infty e^{-u/\xi} 
	{\d u \over u^2} \Biggr (-9 u \csch^2 u - 4 u^3 \csch^2 u
\nonumber \\
& & ~~~
\label{eq:x2int}
	+ 2 u^2 \coth u + 15 \coth u + \frac{8}{15} u^4 \coth u - 6 u^2
\coth^3 u \Biggr ) 
\ea
and we have defined 
\be
\xi = {1 \over B_k} \sqrt{I \over 2} \rmmat{~and~}
u ={ \zeta \over B_k} \sqrt{I \over 2}.
\label{eq:xidef}
\ee
Note that $\xi$ is a dimensionless measure of the strength of the 
field.

The auxiliary functions $X_i$ may be calculated analyically:
\ba
X_0(x) & = & 4 \int_0^{x/2-1} \ln(\Gamma(v+1)) \d v
+ \frac{1}{3} \ln \left ( \frac{1}{x} \right )
+ 2 \ln 4\pi - 4 \ln A-\frac{5}{3} \ln 2 \nonumber \\
& & ~~ - \left [ \ln 4\pi + 1 +  \ln \left ( \frac{1}{x} \right ) \right ] x
+ \left [ \frac{3}{4} + \frac{1}{2} \ln \left ( \frac{2}{x} \right )
\right ]
x^2
\label{eq:x0anal} \\
X_1(x) & = & - 2 X_0(x) + x X_0^{(1)}(x) + \frac{2}{3} X_0^{(2)} (x) -
\frac{2}{9} \frac{1}{x^2}
\label{eq:x1anal} \\
X_2(x) & = & -24 X_0(x) + 9 x X_0^{(1)}(x) + (8 + 3 x^2) X_0^{(2)}(x)
+ 4 x X_0^{(3)}(x) \nonumber \\
& & ~~ - \frac{8}{15} X_0^{(4)}(x) + \frac{8}{15}
\frac{1}{x^2} + \frac{16}{15} \frac{1}{x^4} 
\label{eq:x2anal}
\ea
where
\be
X_0^{(n)}(x) = \frac{\d^n X_0(x)}{\d x^n}
\ee
Because of the near symmetry between $I$ and $K$ in
\eqref{lag1def}, higher derivatives with respect to $K$ may
be calculated in principle, and represented by a sum
of derivatives of $\lag(I,0)$ with respect to $I$.
The constant $A$ is defined as
\ba
\ln A &=& \lim_{n\rarrow\infty} \left ( \sum_{i=1}^n i \ln i \right ) -
\left [ \left ( \frac{n^2}{2} + \frac{n}{2} + \frac{1}{12} \right )
\ln n - \frac{n^2}{4} \right ] \\
&=& \frac{1}{12} - \zeta^{(1)}(-1) = 0.248754477,
\label{eq:lnAdef}
\ea
in analogy to the Euler-Mascheroni constant\cite{Barn00}.  Here
$\zeta^{(1)}(x)$ denotes the first derivative of the
Riemann Zeta function.

Barnes\cite{Barn00} evaluates the definite integral of $\ln \Gamma(x)$ in
terms of the $G$-function
\be
\int_0^{x/2-1} \ln(\Gamma(v+1)) \d v = \left ( \frac{x}{2} - 1 \right
) \ln \Gamma \left ( \frac{x}{2} \right ) - \ln G \left ( \frac{x}{2}
\right )
- \frac{x^2}{8} + \frac{x}{4} \left ( 1 + \ln 2\pi \right ) -
\frac{1}{2} \ln 2 \pi.
\label{eq:Gdef}
\ee
where
\be
G(z) = (2 \pi)^{\frac{z-1}{2}} e^{-\frac{z(z-1)}{2}} e^{-\gamma
\frac{(z-1)^2}{2}} \prod_{k=1}^\infty \left [ \left ( 1 + \frac{z-1}{k}
\right )^k e^{1-z+\frac{(z-1)^2}{2k}} \right ].
\label{eq:Ganal}
\ee
The integral of $\ln \Gamma(x)$ may also be expressed in terms of the
generalized Riemann Zeta function\cite{Ditt79}
\ba
\int_0^{x/2-1} \ln(\Gamma(v+1)) \d v &=&
\zeta^{(1)}\left(-1,\frac{x}{2}-1\right)
- \zeta^{(1)}(-1)
+ \frac{\ln2\pi}{2} \left (\frac{x}{2}-1\right)
\nonumber \\*
& & ~~~
- \frac{x}{4}\left(\frac{x}{2}-1\right)
+ \left (\frac{x}{2}-1\right) \ln \left (\frac{x}{2}-1\right)
\ea
Our expression for $X_0$ was also found by Dittrich \etal\cite{Ditt79}.  
Ivanov\cite{Ivan92} derived a similar expression as well, but his
result differs
from ours and that of Dittrich \etal in the constant term.  Unlike Ivanov's
expression, ours approaches zero as $\xi$ goes to 0 which
from examination of \eqref{x0int} is the correct
limiting behavior.  In addition, the above form for $X_0$ reproduces
the asymptotic strong-field limit given by Heisenberg and Euler\cite{Heis36}.

These functions can be expanded in both the weak-field and
strong-field limits.  In the weak-field limit ($\xi < 0.5$) we obtain
\ba
X_0\left(\frac{1}{\xi}\right) &=&
- \sum_{j=1}^\infty \frac{2^{2j} B_{2(j+1)}}{j(j+1)(2j+1)} \xi^{2j}
\label{eq:X0weak} \\
X_1\left(\frac{1}{\xi}\right) &=&
-\frac{14}{45} \xi^2
+ \frac{1}{3} \sum_{j=2}^\infty \frac{2^{2j} \left (6 B_{2(j+1)} -
(2j+1) B_{2j} \right )}{j(2j+1)} \xi^{2j}
\label{eq:X1weak} \\
X_2\left(\frac{1}{\xi}\right) &=&
\frac{1}{15} \sum_{j=3}^\infty \frac{2^{2j}}{j} \Biggr [
2j (2j-1) B_{2(j-1)} + 60(j-1) B_{2j} \nonumber \\*
& & ~~~~~~~~ - 180 \frac{j-2}{2j+1} B_{2(j+1)}
\Biggr ] \xi^{2j} 
\label{eq:X2weak}
\ea
where $B_j$ denotes the $j$th Bernoulli number.
In the strong-field limit ($\xi > 0.5$), we obtain
\ba
X_0\left(\frac{1}{\xi}\right) &=&
\left ( \frac{1}{3} \ln 2 - 4 \ln A + \frac{1}{3} \ln \xi \right )
+ \left ( 1 - \ln \pi + \ln \xi \right) \xi^{-1}
\nonumber \\*
& & ~~
+ \left ( \frac{1}{2} (\ln 2 - \gamma) + \frac{3}{4} + \frac{1}{2} \ln
\xi \right ) \xi^{-2} \nonumber \\*
& & ~~
+ \sum_{j=3}^\infty \frac{(-1)^{j-1}}{2^{j-2}}
\frac{1}{j} \frac{1}{j-1} \zeta(j-1) \xi^{-j}
\label{eq:X0strong} \\
X_1\left(\frac{1}{\xi}\right) &=&
-\frac{2}{3} \xi
+ \left ( 8 \ln A - \frac{1}{3} - \frac{2}{3} \gamma \right ) 
\nonumber \\*
& & ~~
+ \left ( \ln\pi + \frac{1}{18} \pi^2 - 2 - \ln \xi \right ) \xi^{-1}
+ \left ( -\frac{1}{2} - \frac{1}{6} \zeta(3) \right ) \xi^{-2} 
\nonumber \\*
& & ~~
+ \sum_{j=3}^\infty \frac{(-1)^{j-1}}{2^{j-2}} 
 \left ( \frac{j-2}{j(j-1)} \zeta(j-1) + \frac{1}{6}
\zeta(j+1) \right ) \xi^{-j}
\label{eq:X1strong} \\
X_2\left(\frac{1}{\xi}\right) &=&
\frac{16}{15} \xi^3
- 4 \xi
+ \left ( -6 - 8 \gamma + \frac{4}{15} \zeta(3) + 96 \ln A \right )
\nonumber \\*
& & ~~
+ \left ( -27 + \pi^2 - 15 \ln\pi - \frac{1}{225} \pi^4 - 15 \ln \xi
\right ) \xi^{-1}
\nonumber \\*
& & ~~
+ \left ( \frac{2}{5}\zeta(5) - 9 - 4\zeta(3) \right ) \xi^{-2}
\nonumber \\*
& & ~~+ \sum_{j=3}^\infty \frac{(-1)^{j-1}}{2^{j-2}} \Biggr [
\frac{3(j+4)(j-2)}{j(j-1)} \zeta(j-1)
\nonumber \\*
& & ~~~~~~~~~~ + (j+2) \left ( \zeta(j+1) -
\frac{j+1}{30} \zeta(j+3) \right ) \Biggr ] \xi^{-j}.
\label{eq:X2strong}
\ea

The Lagrangian may also be expanded in terms of the invariants
themselves or the electric and magnetic fields.  In the weak-field
limit we obtain
\ba
\lag_1(I,0)
	 & = & \frac{e^2}{h c} \left ( \frac{1}{180} \frac{I^2}{B_k^2}
- \frac{1}{630} \frac{I^3}{B_k^4} + \frac{1}{630} \frac{I^4}{B_k^6} +
\cdots \right )
\label{eq:X0weaki} \\
\left . \pp{\lag_1}{K} \right |_{K=0} 
	& = & {e^2 \over h c} \left ( -\frac{7}{720} \frac{1}{B_k^2}
 + \frac{13}{5040}
\frac{I}{B_k^4} - \frac{11}{3780} \frac{I}{B_k^6} + \cdots \right )
\label{eq:X1weaki} \\
 \left . \frac{\partial^2 \lag_1}{\partial K^2} \right |_{K=0} 
	& = &  {e^2 \over h c} \left (
\frac{19}{15120} \frac{1}{B_k^6} - \frac{127}{23760} \frac{I}{B_k^8}
+ \frac{5527}{180180} \frac{I^2}{B_k^{10}} + \cdots \right ).
\label{eq:X2weaki}
\ea
This weak-field expansion agrees with the Heisenberg and
Euler\cite{Heis36} result.

In the strong-field limit, for direct comparison with Heisenberg
and Euler\cite{Heis36}
we define $a=E/E_k$ and $b=B/B_k$ and take the limit
$b \gg 1$ and $a \ll 1$.  We take, $\xi^2 = b^2 - a^2$, $\xi \approx b -
\frac{a^2}{2b}$ and $K = - 16 B_k^4 (a b)^2$.  We obtain
\ba
\lag_1(a,b) & = & 4 {e^2 \over h c} B_k^2 \Biggr [
b^2 \left ( \frac{\ln b}{12} - \ln A + \frac{\ln 2}{12} \right ) 
+ \frac{b}{4} \left (\ln b + 1 - \ln \pi \right)  \nonumber \\*
& & ~~~ + \frac{\ln b}{8} + \frac{3}{16} + \frac{\ln 2 - \gamma}{8}
 \nonumber \\*
& & ~~~
- \frac{a^2}{12} \left ( \ln b + \ln 2 - \gamma  \right ) 
\nonumber \\*
& & ~~~ + b \left ( \frac{a^2}{12} + \frac{a^4}{90} + \cdots \right )
+ \cdots
\Biggr ]
\label{eq:lagstrong} 
\ea
which agrees numerically with the corresponding expansion in
Heisenberg and Euler\cite{Heis36}.

\section{Pair Production in an Arbitrarily Strong Electric Field}

In a strong electric field with no magnetic field, the value of
the first invariant is
negative, $I=-2 |\efi|^2$ and $K=0$.  The analytic expressions for the
Lagrangian are valid for values of $\xi$ throughout the complex
plane, with a branch cut along the negative real axis.  Using an
imaginary value of
\be
\xi = i \left ( \frac{E}{E_k} \right ) = i y,~y>0
\label{eq:ydef}
\ee
and
taking $w=2 (4\pi \hbar)^{-1} \IM \lag$ gives the pair
production rate per unit volume \cite{Bere82}.  From examination of
\eqref{X0weak}, for $\xi < 0.5$ the pair-production rate is
apparently zero.  However, since \eqref{X0weak} is a power
series in $\xi/2$, the imaginary part of $X_0$ may be exponentially
small.  Berestetskii \etal\cite{Bere82} derive for a weak field,
\be
w = 2 (4 \pi \hbar)^{-1} \IM \lag \sim  \frac{1}{4 \pi^3}
\left ( \frac{\hbar}{m c} \right )^{-3}
\left ( \frac{\hbar}{m c^2} \right )^{-1}
y^2 \exp \left ( - \frac{\pi}{y} \right )
\ee
To simplify the numerics we use an alternate
definition of $X_0(x)$ obtained by means of a change of variables
\ba
X_0(x) & = & 4 \left ( \frac{x}{2} - 1 \right ) \int_0^1 \ln\left(\Gamma\left(u\left(\frac{x}{2}-1\right)+1\right)\right) \d u
+ \frac{1}{3} \ln \left ( \frac{1}{x} \right )
 \nonumber \\*
& & ~~ + 2 \ln 4\pi  - (4 \ln A+\frac{5}{3} \ln 2) - \left [ \ln 4\pi + 1 +  \ln \left ( \frac{1}{x} \right ) \right ] x
 \nonumber \\*
& & ~~ + \left [ \frac{3}{4} + \frac{1}{2} \ln \left ( \frac{2}{x} \right )
\right ]
x^2.
\label{eq:X0new} 
\ea
With this definition and the property of the Gamma function, $\ln
\Gamma({\bar x}) = \overline{ \ln \Gamma(x)}$, we see that
$X_0({\bar x})=\overline{X_0(x)}$, so
\ba
w & = & 2 (4 \pi \hbar)^{-1} \IM \left . \lag \right |_{I =
-2 y^2 E_k^2, K=0}  \nonumber \\*
&=& 
i \frac{e^2}{8 \pi^2 \hbar^2 c} E_k^2 y^2 \left ( X_0 \left (-\frac{i}{y} \right )
-  X_0 \left (\frac{i}{y} \right ) \right ) \nonumber \\*
&=&
\frac{1}{8\pi^2} \left ( \frac{\hbar}{m c} \right )^{-3} \left (
\frac{\hbar}{m c^2} \right )^{-1} \Biggr [ - \frac{1}{3} \pi y^2
- 8 y^2 \IM Q(y) \nonumber \\*
& & ~~~ 
- 2 y (\ln y + \ln 4\pi + 1) 
+ 4 y \RE Q(y)
+ \frac{1}{2} \pi \Biggr ]
\label{eq:pprate}
\ea
where
\be
Q(y) = \int_0^1 \ln\left(\Gamma\left(u\left(\frac{i}{2y}-1\right)+1\right)\right) \d u
\label{eq:qdef}
\ee
and the scaling constant
\be
\frac{1}{8\pi^2} \left ( \frac{\hbar}{m c} \right )^{-3} \left (
\frac{\hbar}{m c^2} \right )^{-1} = 1.7 \times 10^{51} \rmmat{cm}^{-3} \rmmat{s}^{-1}.
\label{eq:scaledef}
\ee
This expression for $w$, the pair-production rate, agrees numerically with
Itzykson and Zuber's results \cite{Itzy80} for an arbitrarily strong
electric field.

In the strong-field limit we use \eqref{X0strong} and
take the imaginary part
\ba
w  &=&
\frac{1}{8 \pi^2} \left ( \frac{\hbar}{m c} \right )^{-3} \left (
\frac{\hbar}{m c^2} \right )^{-1} \Biggr [
-\frac{\pi}{3} y^2
+ 2 (1 - \ln \pi + \ln y) y
\nonumber \\*
& & ~~~
+ \frac{\pi}{2} + \sum_{k=1}^\infty \frac{(-1)^{3k}\zeta(2 k) }{2^{2k-1}k(2k+1)}
 y^{-(2k-1)}
\label{eq:ppstrong}
\ea

\begin{figure}
\plottwo{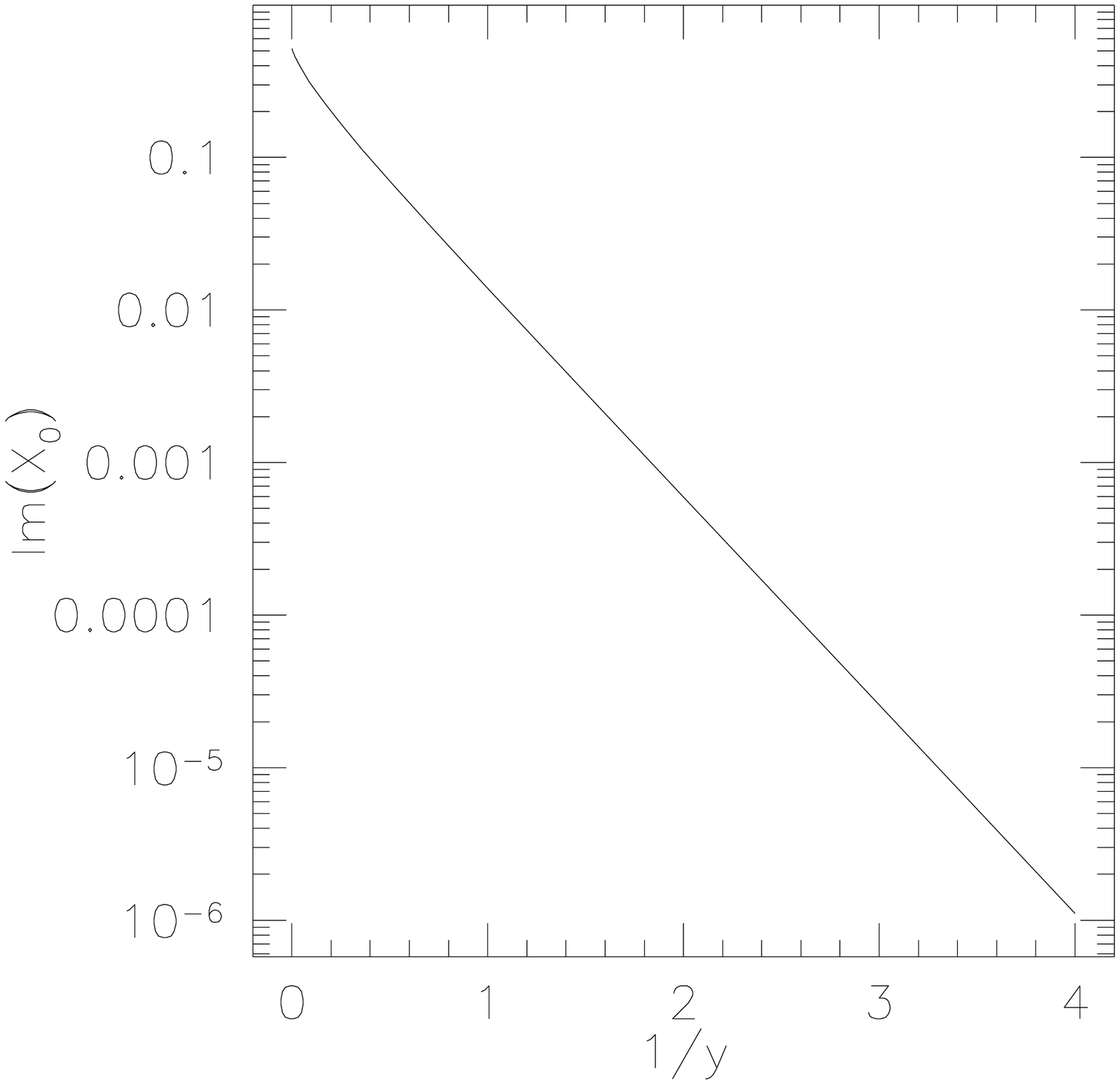}{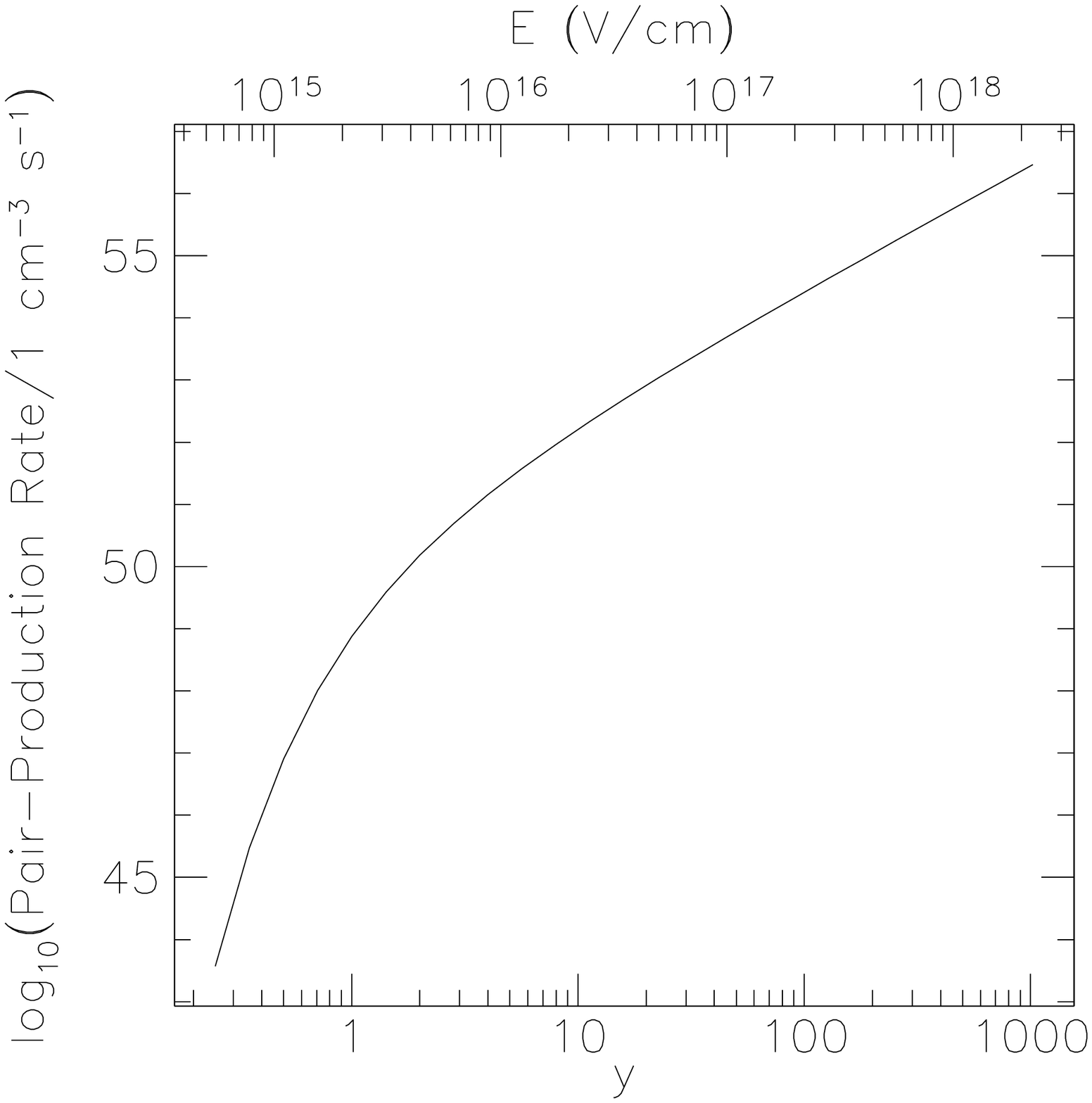}
\caption{The left panel depicts the imaginary part of $X_0$ as a
function of $1/y$.  For weak fields the imaginary component is
approximately $\pi^{-1}\exp(-\pi/y)$.  The right panel depicts the
pair-production rate for near and super-critical fields.}
\label{fig:pprate}
\end{figure}
\figref{pprate} depicts the imaginary component of $X_0$ for
$I=-2 y^2 E_k^2$ and the pair-production rate per unit volume.  From the left
panel, we verify that the imaginary component of $X_0$ is
approximately $\pi^{-1} \exp(-\pi/y)$ for weak fields.
The right panel shows the pair-production rate
which increases as $y^2$ for strong fields and is damped exponentially
in weak fields.

\section{Auxiliary Functions for Photon Splitting}

To calculate the photon splitting rates we follow the technique by
Adler\cite{Adle71}  for the low-frequency limit.
In this limit, Adler expresses
the opacity for photon spliting by means of auxiliary functions which
are simply derivatives of the Lagrangian and therefore of
the functions $X_i$ above
\ba
\kappa \left [ \| \rarrow \|\,+\,\| \right ]  
  & = & \frac{\alpha^6}{2 \pi^2} \frac{\hbar^{13}}{m^{16} c^{24}}
B^6 \sin^6 \theta \frac{\omega^5}{30} C_1(\xi)^2 \\
  & = & \frac{\alpha^3}{2 \pi^2} \left ( \frac{B \sin\theta}{B_k} \right)^6
\left( \frac{\hbar \omega}{m c^2} \right )^5 \frac{C_1(\xi)^2}{30}
 \left ( \frac{m c}{\hbar} \right ) \\
  & = & 17.0 \rmmat{\,cm}^{-1} \left ( \frac{B \sin\theta}{B_k} \right)^6
\left( \frac{\hbar \omega}{m c^2} \right )^5 C_1(\xi)^2 
\\
\kappa \left [ \| \rarrow \perp + \perp \right ]  
  & = & \frac{\alpha^6}{2 \pi^2} \frac{\hbar^{13}}{m^{16} c^{24}}
B^6 \sin^6 \theta \frac{\omega^5}{30} C_2(\xi)^2 \\
\kappa \left [ \perp \rarrow \|\, + \perp \right ] 
  & = & 2 \frac{\alpha^6}{2 \pi^2} \frac{\hbar^{13}}{m^{16} c^{24}}
B^6 \sin^6 \theta \frac{\omega^5}{30} C_2(\xi)^2,
\ea
The conversion of $\perp$ to $\|\, + \perp$ proceeds through two
channels hence the two-fold increase in the opacity for this process.
$C_1$ and $C_2$ are defined by
\ba
\left . \frac{\partial^3 \lag_\rmscr{Adler}}{\partial {\cal F}^3} \right |_{{\cal
G}=0,{\cal F} = \frac{1}{2} B^2} & = & \frac{64}{4 \pi} \left
. \frac{\partial^3 \lag_1}{\partial I^3} \right |_{
K=0,I = 2 B^2}  \\
& = & - \frac{\alpha^3 \hbar^6}{2 \pi^2 m^8 c^{10}} C_1(\xi)
= - \frac{\alpha}{2 \pi^2 B_k^4} C_1(\xi)\\
\left . \frac{\partial^3 \lag_\rmscr{Adler}}{\partial {\cal F} \partial {\cal G}^2} \right |_{{\cal
 G}=0,{\cal F} = \frac{1}{2} B^2} & = & -\frac{128}{4 \pi} \left
. \frac{\partial^2 \lag_1}{\partial I \partial K} \right |_{
K=0,I = 2 B^2}  \\
& = & -  \frac{\alpha}{2 \pi^2 B_k^4} C_2(\xi)
\ea
and $\theta$ is the angle between the direction of propagation of the
photon and the external magnetic field.  
The factors of 128 and 64 result from the definitions of
Adler's ${\cal F}$ and ${\cal G}$ in terms of $I$ and $K$,
\ba
{\cal F} &=& \frac{1}{2} \left ( |\bfi|^2 - |\efi|^2 \right ) =
\frac{I}{4} \\
{\cal G} &=& \bfi \cdot \efi = \frac{1}{4} \sqrt{-K}
\ea
An additional factor of $4 \pi$ appears because we are using
rationalized Gaussian units while Adler employs unrationalized units.

Given the analytic forms for $X_0$ and $X_1$ we obtain
\ba
C_1(\xi) &=& \frac{1}{4 \xi^7} \left (
X_0^{(3)}\left(\frac{1}{\xi}\right)
+ 3 X_0^{(2)}\left(\frac{1}{\xi}\right) \xi
- 3 X_0^{(1)}\left(\frac{1}{\xi}\right) \xi^2 \right ) \\
C_2(\xi) &=& -\frac{1}{4 \xi^5} \left (
X_1^{(1)}\left(\frac{1}{\xi}\right)
+ 2 X_1\left(\frac{1}{\xi}\right) \xi \right ) 
\ea
The functions, $C_1$ and $C_2$, have the
appropriate limits as $\xi\rarrow 0$ which
correspond to the lowest order hexagon diagrams for the splitting
process,
\ba
C_1(I) &=& 16 \sum_{j=2}^\infty \frac{2^{j-1} (j-1) B_{2(j+1)}}{2j+1}
\left ( \frac{I}{B_k^2} \right )^{j-2} \\
       &=& 6 \cdot \frac{8}{315}
- \frac{64}{105}  \frac{I}{B_k^2}
+ \frac{320}{99} \left ( \frac{I}{B_k^2} \right )^2 \cdots \\
C_2(I) &=& \frac{1}{3} \sum_{j=2}^\infty \frac{2^{j+1} (j-1)
\left ( 6 B_{2(j+1)} - (2j+1) B_{2j} \right )}{j(2j+1)}  
\left ( \frac{I}{B_k^2} \right )^{j-2} \\
       &=& 6 \cdot \frac{13}{945} 
- \frac{176}{945}  \frac{I}{B_k^2}
+ \frac{332}{495} \left ( \frac{I}{B_k^2} \right )^2 \cdots 
\ea
In the strong field limit we obtain
\ba
C_1(\xi) &=&
\frac{1}{3} \xi^{-4}
- \frac{3}{4} \xi^{-5} \left (\ln \xi - \ln \pi +\frac{2}{3} \right )
- \xi^{-6} \nonumber \\*
& & ~~~
- \sum_{j=3}^\infty \frac{(-1)^j}{2^j}
 \frac{j^2-4}{j-1} \zeta(j-1) \xi^{-j-4}  \\
C_2(\xi) &=&
\frac{1}{6} \xi^{-3}
+ \xi^{-4} \left (\frac{1}{6} + \frac{1}{3} \gamma - 4 \ln A \right)
\nonumber \\*
& & ~~~
+ \frac{3}{4} \xi^{-5} \left (\ln \xi - \ln \pi + \frac{5}{3} -
\frac{1}{18}\pi^2 \right ) + \frac{1}{2} \xi^{-6} \left ( 1 + \frac{1}{3} \zeta(3) \right)
\nonumber \\*
& & ~~~
+  \sum_{j=3}^\infty \frac{(-1)^{j}(j+2)}{2^j} 
 \left ( \frac{j-2}{j(j-1)} \zeta(j-1) + \frac{\zeta(j+1)}{6}
\right ) \xi^{-j-4}
\ea
These expressions for the photon-splitting rate are only valid in the
low-frequency limit since the Heisenberg-Euler Lagrangian neglects
the gradients of the field.   When these gradients are neglected,
the results from Schwinger's proper-time integration\cite{Schw51}
used by Adler\cite{Adle71} reduce to these results
obtained from the Heisenberg-Euler Lagrangian.  Ba{\u\i}er, Milstein and
Shaisultanov \cite{Baie96} have also obtained similar results for
arbitrary field strengths (and photon frequencies) using an operator
diagram technique.

Ba{\u\i}er \etal and Adler's methods differ in spirit but yield
the same results.  Ba{\u\i}er, Katkov, and Strakhovenko developed the
operator diagram technique \cite{Baie75a}.  In this formalism, the
photon splitting matrix element is evaluated with Feynman diagrams
\cite{Baie75b} using electron propagators in an external field.  On
the other hand, Adler\cite{Adle71} calculates the expectation value of
the current ($\left<j_\mu(x)\right>$) order by order in the external
photon fields using Schwinger's formalism \cite{Schw51} and relates
this expectation value to the photon-splitting matrix element.  

\section{Photon Splitting Opacities and Application to Neutron Stars}

Adler \cite{Adle71} argues that because of dispersive effects, the process 
$\| \rarrow \perp + \perp$, dominates the opacity of photons
travelling through a strong field.  Therefore, we are interested in
the function $C_2(\xi)$ which determines the splitting rate for all
magnetic field strengths at photon energies small compared to the mass
of the electron.  We see immediately from the expansions of $C_2$ that
the opacity has the following behavior for weak and strong fields
\be
\kappa \left [ \| \rarrow \perp + \perp \right ]  =
\left \{ \begin{array}{ll}
0.116 \rmmat{\,cm}^{-1} \left ( \frac{B \sin\theta}{B_k} \right)^6
\left( \frac{\hbar \omega}{m c^2} \right )^5 & B \ll B_k \\*
\\*
0.472 \rmmat{\,cm}^{-1} \sin^6 \theta \left( \frac{\hbar \omega}{m c^2} \right )^5 &
B \gg B_k
\end{array}
\right .
\ee
We find, in agreement with the recent result of Ba{\u\i}er \etal
\cite{Baie96} and as well as with earlier results \cite{Thom95,Bari95a}
that the 
photon splitting opacity approaches a constant value in the limit of
strong fields.

The left panel of \figref{kappa} depicts the opacity for
photons with $E=mc^2$ as a function of $\xi$.  Our formulae are not
valid for these high-energy photons but for low energies the opacity
scales as this quantity times the photon energy to the fifth power.
The right panel applies these opacities to neutron stars.  Neutron
stars are observed to have magnetic fields $\sim 10^{12}$\,G
(\eg\ \cite{Shap83}) and a subset of these objects known as magnetars are
suspected to have much stronger fields $\sim 10^{16}$\,G or larger
\cite{Dunc92}.  The figure illustrates the energy of photons with a
mean-free path of ten kilometers.  All parallel-polarized photons
with this energy or larger would tranverse an optical depth of one or
larger while escaping from the neutron star.
\begin{figure}
\plottwo{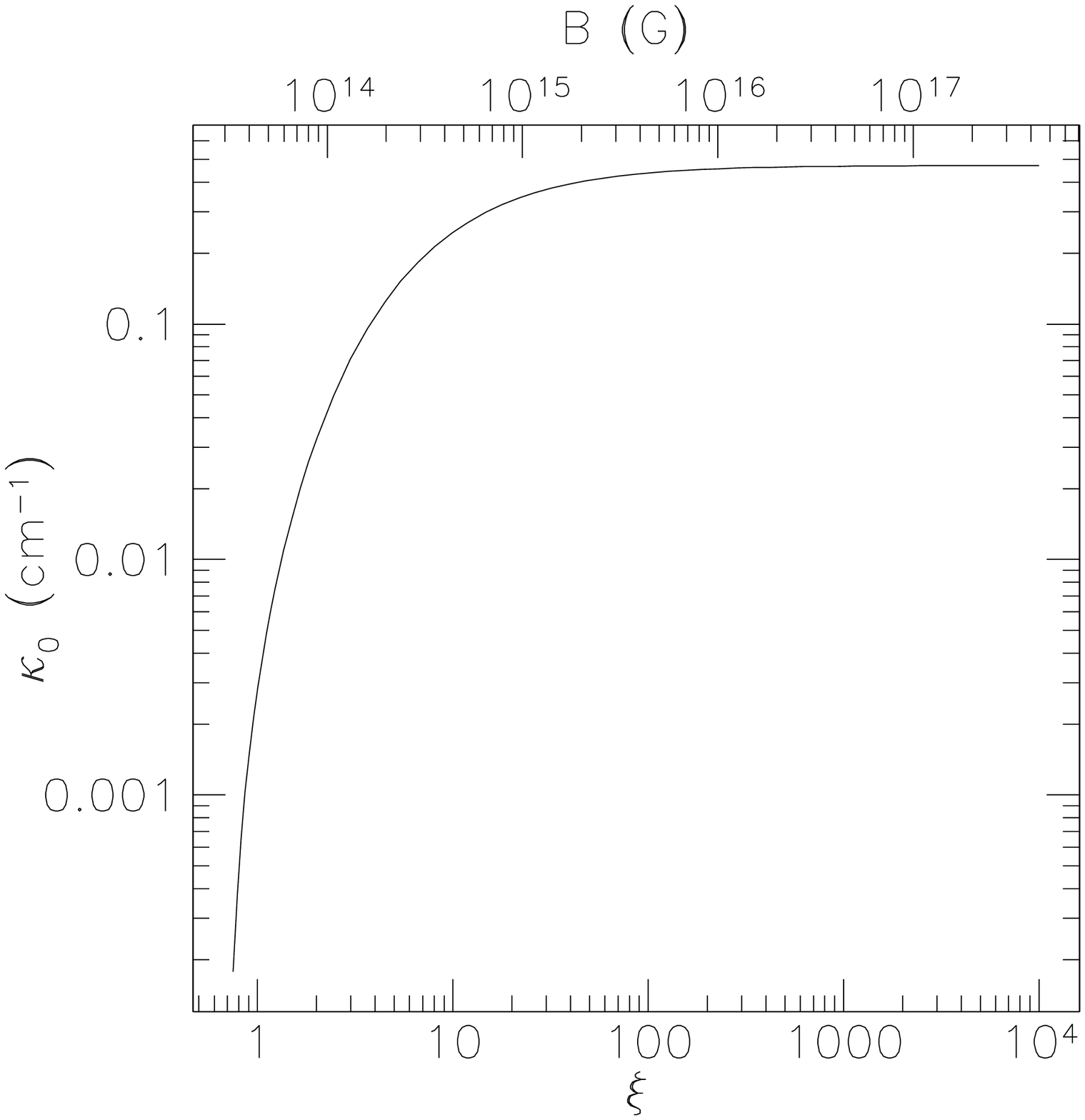}{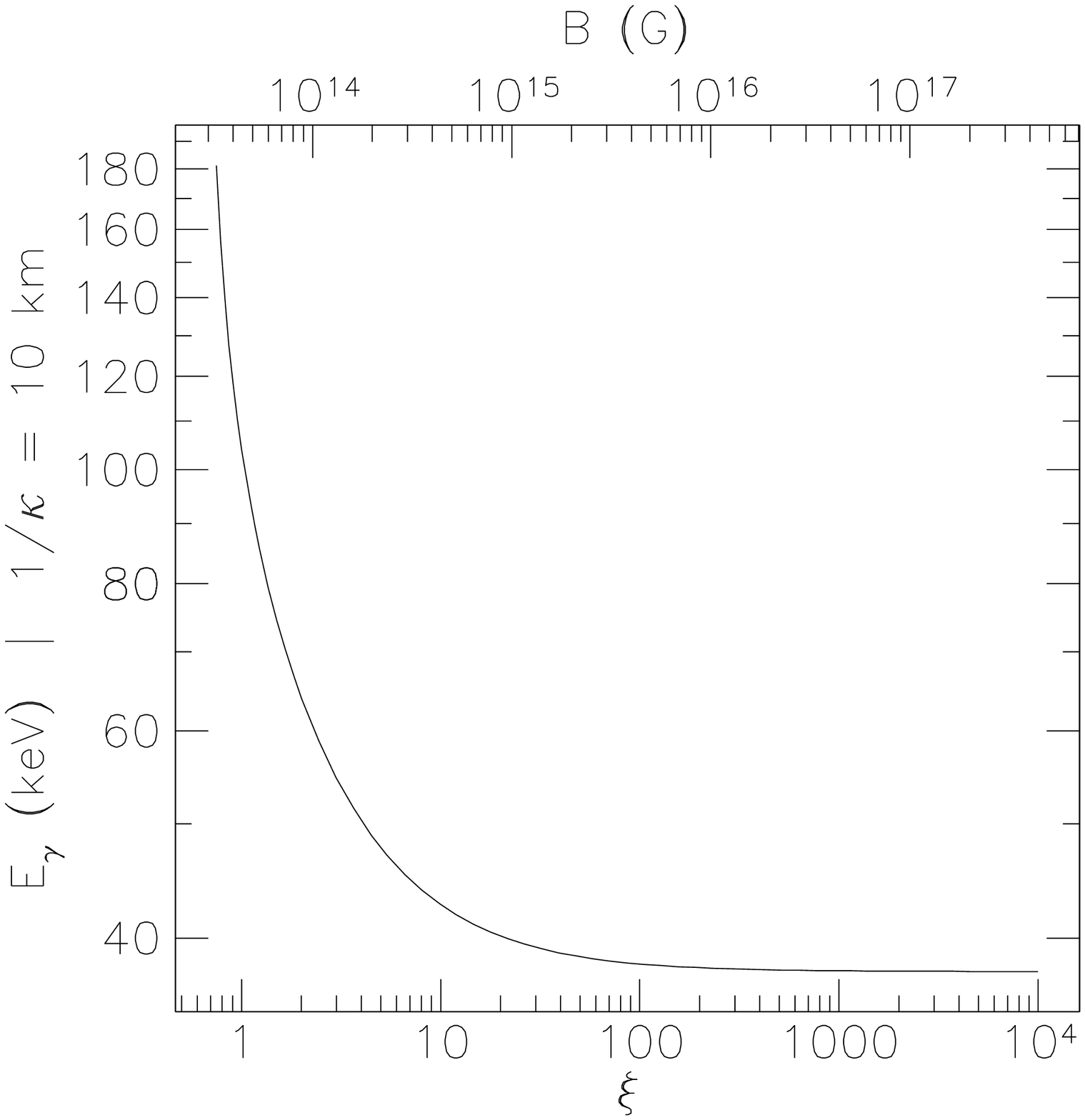}
\caption{The left panel depicts the photon-splitting opacity for
photons with $E=mc^2$ as a function $\xi$ and $B$ (upper axis).
The right panel shows the energy of photons
with a mean-free path for splitting of ten kilometers as a function
$\xi$ and $B$}
\label{fig:kappa}
\end{figure}

\section{Conclusion}

Because of the asymptotic behavior of the function $C_2(\xi)$, even in
immensely large fields, the photons with energies less than $37$ keV
have opacities less than (10 km)$^{-1}$.  This energy corresponds to a
temperature of $4 \times 10^8$ K, so we must conclude that unless the
strong magnetic field of the neutron star extends over a distance much
greater than $10$ km, photon splitting affects the thermal radiation
of only the youngest neutron stars.

We have derived a closed form expression for the Heisenberg-Euler
effective Lagrangian for quantum electrodynamics as a function of the
gauge and Lorentz invariant quantities $I$ and $K$ in the limit of
small $K$.  We have calculated from this analytic expression the
photon-splitting and pair-production rates in the intense field and
found them to agree with previous work.  Furthermore, the
expressions for the dielectric and permeability tensors in an external
field derived from our analytic expression also agree with previous
results \cite{Heyl96b}.
We expect that these expressions may be applied to a wide
variety of problems in strong electromagnetic fields, including
Compton scattering, photon-photon scattering, and bremsstrahlung.

\acknowledgements

This material is based upon work supported under a National Science
Foundation Graduate Fellowship.  L.H. thanks the National Science
Foundation for support under the Presidential Faculty Fellows Program.
We would also like to thank the anonymous referee for many useful comments.

\bibliography{qed,ns,math}
\bibliographystyle{prsty}

\end{document}